\documentclass[aps,prb,twocolumn]{revtex4}

\usepackage{times}
\usepackage{graphicx}
\usepackage{float}
\usepackage{latexsym,amsmath,amssymb,bm,euscript}
\usepackage{color}
\usepackage{subfigure}
\usepackage{epstopdf}
\usepackage[colorlinks=true,linkcolor=blue,citecolor=blue]{hyperref}
\usepackage{type1cm}
\fontsize{12pt}{18pt}

\usepackage{appendix}

\begin{document}
\title{Interaction-driven fractional quantum Hall  state of hard-core bosons on kagome lattice at one-third filling}
\author{W. Zhu$^1$, S. S. Gong$^{2}$ and D. N. Sheng$^1$}
\affiliation{$^1$Department of Physics and Astronomy, California State University, Northridge, California 91330, USA}
\affiliation{$^2$National High Magnetic Field Laboratory, Florida State University, Tallahassee, Florida 32310, USA}

\begin{abstract}
There has been a growing interest in realizing topologically nontrivial
states of matter in band insulators, 
where a quantum Hall effect can appear as an intrinsic property of the band structure.
While the on-going progress is under way with a number of directions, 
the possibility of realizing novel interaction-generated topological phases, 
without the requirement of a nontrivial invariant encoded in single-particle wavefunction or band structure, 
can  significantly extend the class of topological materials and is thus of great importance.
Here, we show an interaction-driven topological phase emerging 
in an extended Bose-Hubbard model on kagome lattice, 
where the non-interacting band structure is topological trivial 
with zero Berry curvature in the Brillouin zone.
By means of an unbiased state-of-the-art density-matrix renormalization group technique,
we identify that the groundstate in a broad parameter region is equivalent to 
a bosonic fractional quantum Hall Laughlin state,
based on the characterization of unverisal properties including 
groundstate degeneracy, edge excitations and anyonic quasiparticle statistics.
Our work paves a way of finding interaction induced topological phase
at the phase boundary of conventionally ordered solid phases.
\end{abstract}


\maketitle


\newpage
\section{Introduction}




While fundamental particles in nature are either bosons or fermions,
the emergent excitations in two-dimensional strongly-correlated systems may
obey fractional or anyonic statistics \cite{Tsui1982,Laughlin1983}.
The most famous example is the quasiparticles or quasiholes in fractional quantum Hall (FQH) effects in strong magnetic field, 
which are  topological states of matter whose low-energy physics is governed by the Chern-Simon gauge theory.
Interestingly, recent theoretical discoveries have revealed that FQH effects can also be  realized 
in lattice systems in the absence of an external magnetic field \cite{Titus2011,Tang2011,KaiSun2011,DNSheng2011,Bernevig2011}.
Such a lattice realization  of FQH phases is attributed to two key points: 
a nearly dispersionless single-particle energy band with non-trivial topology characterized by a nonzero Chern number and
a strong many-body interaction comparing to the band width.
The condition of nearly flat topological band is esentially important,
as  the kinetic energy of particles can be quenched in such a topological band akin to the Landau level physics.
The strong interaction also plays the vital role in stabilizing the FQH phases.
In fact, without the interaction or interaction being a subleading correction,
the system is expected to be in a Fermi liquid like state at fractional fillings rather than forming a topological phase.

Given  above facts, a natural question that we address in this paper
is whether a FQH phase is also possible in a lattice model with trivial non-interacting band.
Actually, there have been a series of proposals along this direction \cite{Wen1989,Wen1991,Raghu2008,WenJun2010,KaiSun2009,YZhang2009}, 
where the common wisdom is that the presence of strong interactions in a strongly frustrated system can give rise to a nonlocal
complex bond order parameter and a spontaneous breaking of time-reversal symmetry (TRS) 
through flux attachment\cite{Wen1989}.
However, lacking of a theoretical method to predict the quantum phase for
microscopic interacting systems, theoretical studies usually resort to different mean-field approximations, 
which often favor topological phases.   
As an example,  Raghu \textit{et. al} \cite{Raghu2008} 
showed that a quantum anomalous Hall effect can be dynamically generated in an extended Hubbard model on the honeycomb lattice. 
The similar idea has been applied to other lattice systems with a quadratic band crossing point, 
such as kagome \cite{WenJun2010}, checkboard \cite{KaiSun2009}, diamond \cite{YZhang2009} and Lieb lattice \cite{Tsai2015}.
However, comprehensive numerical studies have been searching for true groundstates in different lattice systems 
and failed to find exotic topological phases predicted by the mean-field theories \cite{Motruk2015,Capponi2015,Ferhat2014,Pollmann2014,Nishimoto2010, SSGong_SQ}.
Here a crucial difficulty is that, instead of triggering the desired TRS spontaneously breaking,
the strong interactions also tend to stabilize competing solid orders by breaking translantional symmetry.
Thus, the simple concept of realizing interaction-induced FQH phases in topological trivial bands 
was  illusive in realistic lattice models.

Very recently, theoretical studies of extended kagome antiferromagnetic systems 
have discovered a particular class of spin liquids,
the so-called chiral spin liquid \cite{anderson1973,Kalmeyer1987}, 
with TRS spontaneously \cite{SSGong_SR,YCHe2014} (or explicitly \cite{Bauer2014}) broken,
which shed lights on this elusive area: 
Long-ranged frustrated interactions may favor the FQH-like ground state 
near the boundary between ordered states \cite{SSGong2015}. 
So far the existing examples are rare and
all occur at the half-filling (half of spins are pointing up in z-direction) on kagome lattice, 
which may be  attributed to quantum fluctuations near the non-coplanar spin ordered state (cuboc phase) \cite{SSGong2015,Wietek2015,Messio2012}. 
Thus it is highly desired to search for the interaction-induced FQH phase
beyond the half-filling, which serves as the proof of the principle that TRS broken
phase can emerge in more general conditions without a nearby cuboc phase. 
On the other hand, kagome-based magnetic systems have been widely studied 
under an external magnetic field \cite{Ishikawa2015}, which can tune the spin systems into different magnetizations
corresponding to hardcore boson systems at different fillings. 
The interesting candidate states  have been established including 
the valence bond crystal state \cite{Isakov2006,Nishimoto2013,Capponi2013} and 
the featureless Mott insulator \cite{Parameswaran2013} as possible groundstates.
Beside  these topological trivial phases, 
 a $Z_2$ topological phase may survive in an easy-axis kagome system \cite{Roychowdhury2015}, 
which is currently  under debate \cite{Plat2015}. 
It is theoretically proposed  that FQH state can also emerge at 1/3 filling \cite{Kumar2014,Kumar2015},
however, so far this possibility has not been established by controlled theoretical methods 
beyond mean-field approaches.
Along this line, the existence of a topological ordered phase at one-third filling remains open.


In this paper, we study an Bose-Hubbard model on the kagome lattice in the hard-core limit:
\begin{eqnarray}
H= &&t\sum_{\langle\mathbf{r}\mathbf{r}^{ \prime}\rangle} \left[b^{\dagger}_{\mathbf{r}^{\prime}}b_{\mathbf{r}}+\textrm{H.c.}\right]
+ V_1 \sum_{\langle\mathbf{r}\mathbf{r}^{ \prime}\rangle} n_{\mathbf{r}} n_{\mathbf{r}^{\prime}}  \nonumber\\ 
&+& V_2 \sum_{\langle\langle\mathbf{r}\mathbf{r}^{ \prime}\rangle\rangle} n_{\mathbf{r}} n_{\mathbf{r}^{\prime}} + 
V_3 \sum_{\langle\langle\langle\mathbf{r}\mathbf{r}^{ \prime}\rangle\rangle\rangle} n_{\mathbf{r}} n_{\mathbf{r}^{\prime}} \label{hamilton}
\end{eqnarray}
, where $b^{\dagger}_{\mathbf{r}}$ ($b_{\mathbf{r}}$) creates (annihilates) a hard-core boson at site $\mathbf{r}$. $t=1$ is the nearest-neighbor hopping amplitude, $V_1, V_2, V_3$ are the density-density repulsion strengths  on first, second and third nearest neighbors, respectively. We focus on the boson filling number $1/3$ in this paper. This model can also be mapped onto the spin$-1/2$ XXZ model, allowing for an interpretation of our results in terms of both bosons and quantum spins. 
The energy band for hosting hardcore bosons is topological trivial (with real hopping terms) with zero Berry curvature in the Brillouin zone.
In order to study the ground state phase diagram in the $\{V_1,V_2,V_3\}$ parameter space, 
we implement the density-matrix renomralization group (DMRG) algorithm on cylinder geometry 
combined with the exact diagonalization (ED) on torus geometry (see Appendix \ref{method} for computational details), 
both of which have been proven to be powerful and complementary tools for studying realistic models 
containing arbitrary strong and frustrated interactions. 


\section{Results}

\subsection{Phase Diagram}
\begin{figure}[t]
\includegraphics[width=0.44\linewidth]{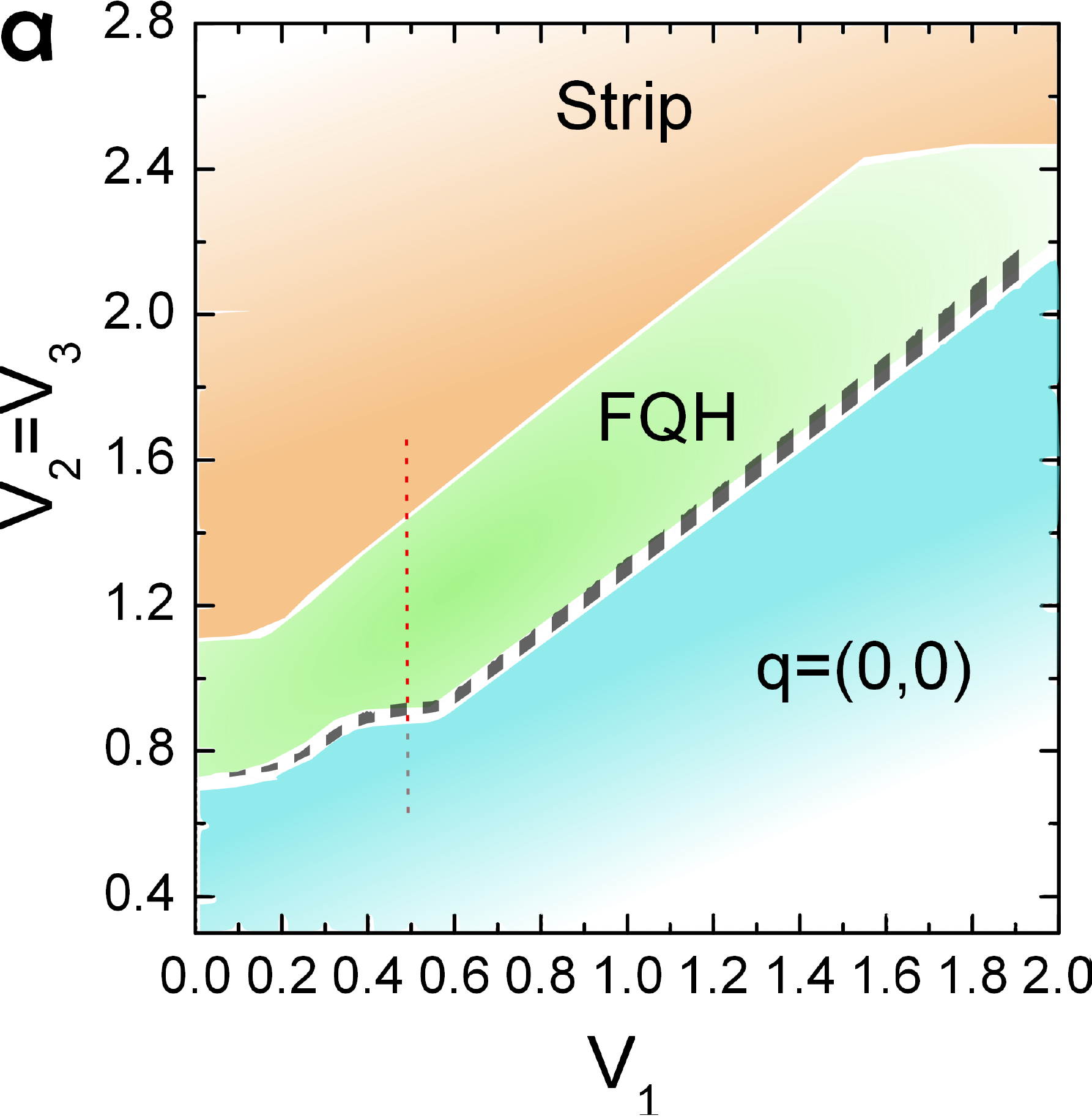}
\includegraphics[width=0.45\linewidth]{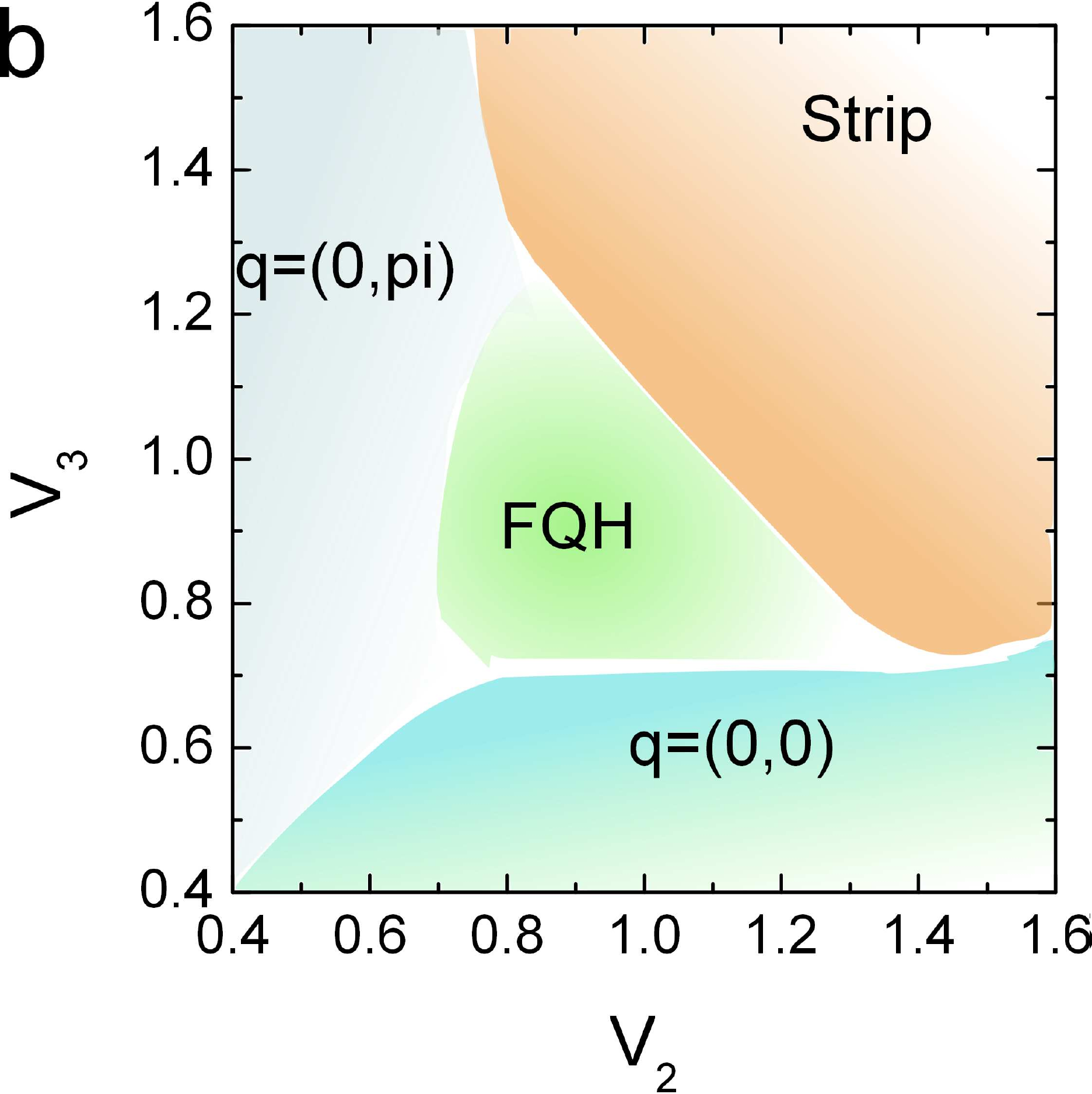}
\includegraphics[width=0.32\linewidth]{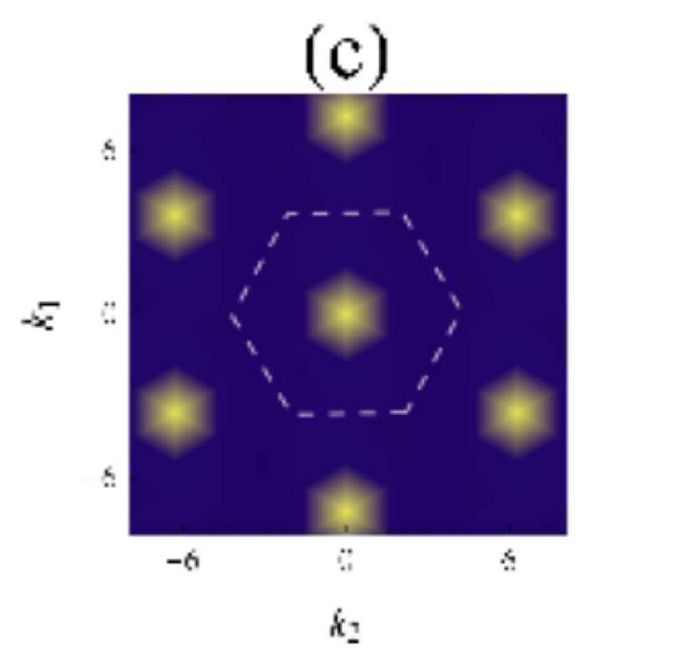}
\includegraphics[width=0.32\linewidth]{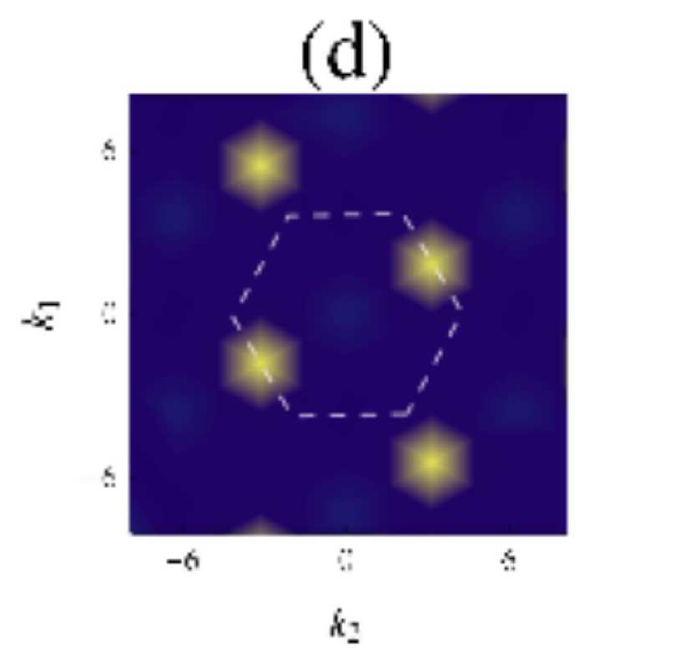}
\includegraphics[width=0.32\linewidth]{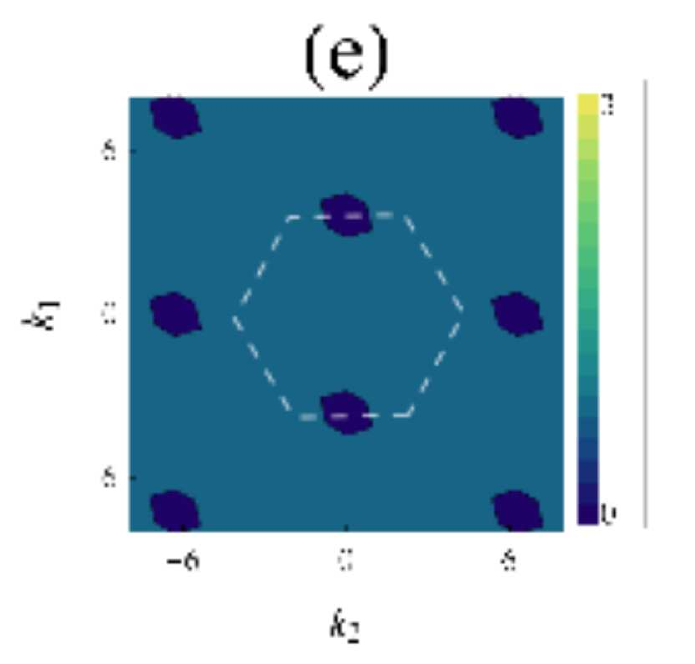}
\caption{Phase diagram of the extended Bose-Hubbard model on kagome lattice 
obtained by DMRG calculations on cylinder of circumference $L_y=4$.
(a)The phase diagram plotted in $V_1$ and $V_2=V_3$ parameter space. The shaded area is a coexistence region.
(b)The phase diagram plotted in $V_2$ and $V_3$ parameter space by setting $V_1=0$.
The contour plots of static density structure factor for: 
(c) charge density wave $q=(0,0)$ phase, (d) strip phase and (e) FQH phase. 
The white dashed line shows the first Brillouin zone. 
}  \label{phase}
\end{figure}

Our main findings are summarized in the phase diagram Fig. \ref{phase}(a-b). 
In the parameter region $V_2\approx V_3$ and $V_2,V_3>V_1$ ($0\leqslant V_1 \leqslant 2.0$),
we find a robust FQH phase emerging with the TRS spontaneously breaking. 
The FQH phase is centered around the line  $V_2\approx V_3$, 
which  persists to $|V_2-V_3|<0.1$ approximately as shown for  $V_1=0$  in Fig. \ref{phase}(b).
The FQH phase is characterized by a four-fold groundstate degeneracy on torus geometry, 
which arises from two sets of Laughlin $\nu=1/2$ FQH states with opposite chiralities.
In particular, the TRS spontaneous breaking can be inspected 
by identifying local circulating currents on the cylinder geometry,
while  the nature of bosonic Laughlin $\nu=1/2$ state will be  identified by
the edge excitation spectrum, fractional Chern number and the anyonic quasiparticle statistics as  elaborated later.
Interestingly,  we also show  that the FQH liquid phase is neighboring with several solid phases which all respect TRS: 
a strip phase, a charge density wave  $q=(0,0)$ phase, and  $q=(0,\pi)$ phase with $q$ as the ordering wave vectors. 
Compared to the distinctive Bragg peaks in the structure factor for solid phases (Fig. \ref{phase}(c-d)), 
the FQH phase shows a structureless feature (Fig. \ref{phase}(e)).
Finally, between the FQH phase and $q=(0,0)$ phase, 
there exists a narrow window for the coexistence of both FQH and charge density order 
(labeled by shaded area in Fig. \ref{phase}(a)). 
By comparing our quantum phase diagram with  the classical phase diagram, we find that the FQH phase
is present near  the classical phase boundary neighboring with  different solid phases (Appendix \ref{classic}).


\subsection{Energy Spectrum and Doubled Topological Degeneracy}
\begin{figure} 
\includegraphics[width=0.99\linewidth]{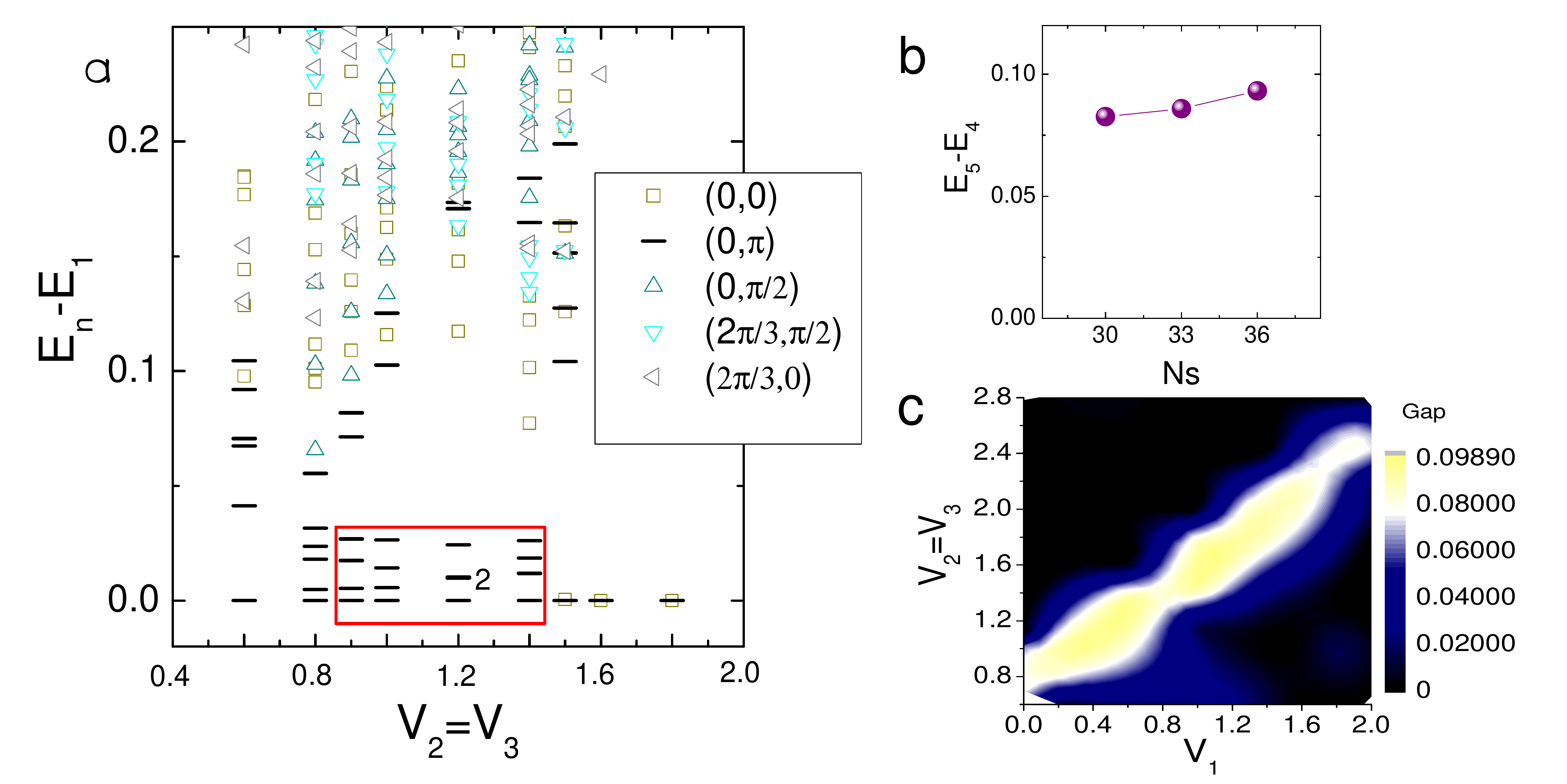}
\caption{Energy spectra from ED calculation. 
(a) A scan across the reference line in Fig. \ref{phase} by setting $V_1=0.5$ on the $N_s=3\times 3\times 4=36$ sites cluster. 
The four-fold degeneracy of the FQH state is present around $0.8< V_2=V_3<1.5$. Different colors and symbols correspond to differeent momentum  sectors. (b) Energy gap for various system sizes $N_s$ by setting $V_1=0.5,V_2=V_3=1.2$.
(c)Contour plot of energy gap versus $V_1$ and $V_2=V_3$ on $N_s=36$ sites cluster.
} \label{ED_energy}
\end{figure}

Topological ordered states have characteristic groundstate degeneracy on compactified space (i.e. torus) while TRS spontaneous
breaking topological phase has doubled topological degeneracy.
To demonstrate this property in the intermediate FQH region, we first investigate the low-energy spectra based on ED calculation. 
In Fig. \ref{ED_energy}, we show  the scan of energy spectra along the line  $V_1=0.5$ and varying $V_2$ ($=V_3$). 
It is clear shown  that there is a 
 fourfold groundstate degeneracy in the regime $0.8<V_2=V_3<1.5$, which is separated by higher
excited states by a robust spectrum gap. 
The fourfold degeneracy arises from
two-fold  topological degeneracy for the  $\nu=1/2$ Laughlin state (full evidences will be shown below)
and an additional  factor of $2$ from two-copies of states with the opposite chirality due to the TRS spontaneously breaking. 
All four groundstates are located in momentum sector $k=(0,0)$, 
consistent  with the expectation of the momentum folding  rule for emerging FQH state
with $12$ particles on $3\times 3\times 4$ kagome lattice \cite{DNSheng2011,Bernevig2011}.
The low-energy spectrum gap for different system sizes is shown in Fig. \ref{ED_energy}(b), where we find that
the energy gap between the fourth lowest energy state and the fifth one is robust against the 
increase of system sizes (Fig. \ref{ED_energy}(b)). This result  indicates the emergent FQH  phase may be robust 
at thermodynamic limit,  which will be further  confirmed by our larger system results 
based on DMRG.         
We also find that the energy gap is robust in the whole FQH regime 
while it drops to near zero at the phase boundary as illustrated in Fig. \ref{ED_energy}(c).

\begin{figure}[!htb]
\includegraphics[width=0.45\textwidth]{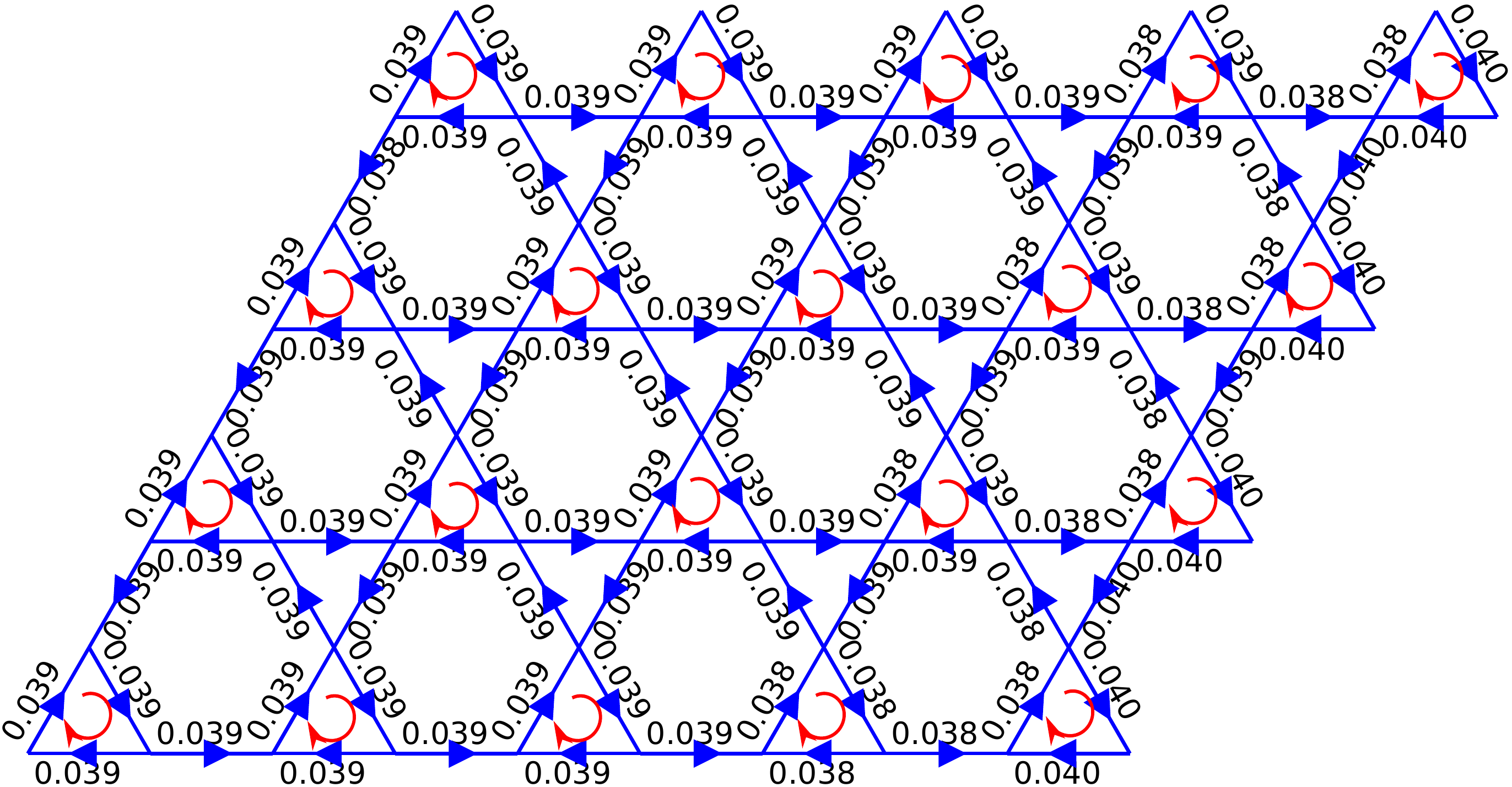}
\caption{Local current $\mathcal{J}_{ij}$ pattern of $|\Psi^{L}_{\openone} \rangle$ 
 from DMRG calculation. 
The parameter is $V_1=0.5,V_2=V_3=1.2$ on $L_y=4$ cylinder (only show the five columns in the center). 
Width of bond is proportional to the current value ($\mathcal{J}_{ij}$ is shown on the bond as a number) and
arrows correspond to current directions.} \label{current}
\end{figure}

\subsection{Time Reversal Symmetry Spontaneously Broken and Local Current}
Since the Hamiltonian (Eq. \ref{hamilton}) preserves TRS, the emergent FQH
state as the groundstate breaks  TRS \textit{spontaneously}.
To investigate this mechanism, we move to larger systems with the  cylinder geometry  and obtain the 
groundstates $|\Psi^{L(R)}_{\openone,s}\rangle$ by implementing DMRG calculation. 
By using different random initial wavefunction in DMRG,  we obtain four different groundstates.
Here we label topologically different groundstates by their chiral and anyonic nature on cylinder geometry, 
where $L$ ($R$) stands for ``left-hand'' (``right-hand'') chirality, 
and $\openone$ ($s$) stands for identity (semion) quasiparticle (see below).) 
A simple picture of the TRS broken can be obtained by measuring  the local circulating currents in real-space. 
As shown in Fig. \ref{current},  the current $\mathcal{J}_{ij}=Im\langle \Psi^{L}_{\openone}|b^{\dagger}_i b_j|\Psi^{L}_{\openone}\rangle$ 
between two nearest-neighbor sites $(i,j)$
forms loop structure and is in the clockwise direction in each triangle in the bulk of the system, 
which is refered as ``left-hand'' chirality. 
The emergent loop current  is a direct demonstration  of TRS spontaneously breaking for the  state,
which enables experimental detecting from local current measurements.   
This is a distinct properties different from the chiral spin liquid at half filling, 
where only the three spin chirality term is nonzero \cite{SSGong_SR,YCHe2014}.
We also find that  the TRS partner $|\Psi^{R}_{\openone,s}\rangle$ hosts anti-clockwise loop current in each triangle. 

\subsection{Fractionalization and Fractional  Statistics for Quasiparticles}
To uncover the anyonic nature of groundstates in the intermediate region, 
we investigate  hallmark signatures  of FQH state including characteristic excitation spectrum on the edge,
fractional Chern number and  quasiparticle braiding statistics in the bulk. 
All the evidences we obtain fully  support the topological phase  in the intermediate region is the 
emerging  bosonic $\nu=1/2$ Laughlin state.

\subsubsection{Entanglement Spectrum}
\begin{figure} 
 \begin{minipage} {0.98\linewidth}
 \centering
 \includegraphics[width=0.95\linewidth]{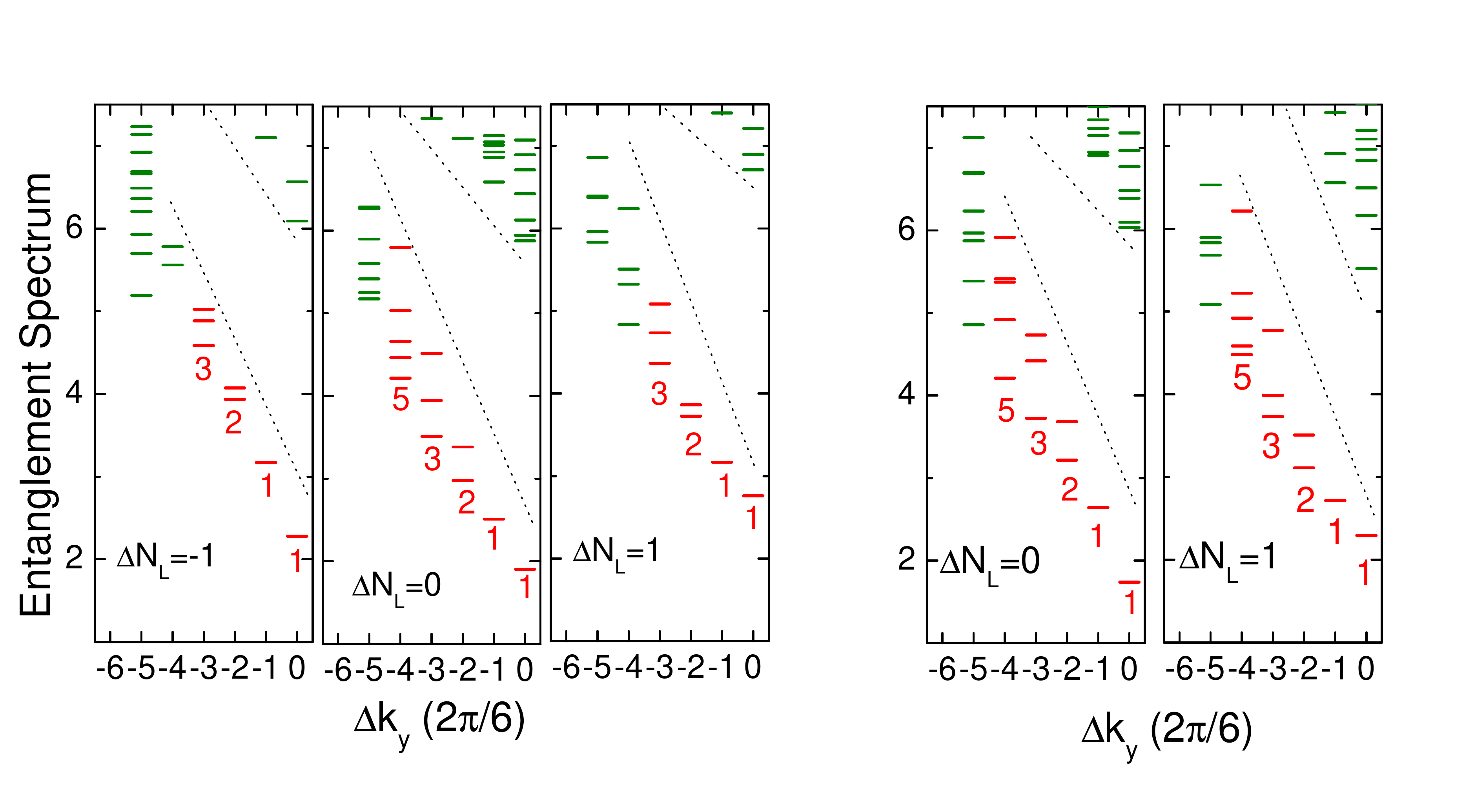}
 \end{minipage}
 \caption{The low-lying ES of $|\Psi^{L}_{\openone}$ (left) and $|\Psi^{L}_{s}$ (right) on $L_y=6$ cylinder.
 The ES is labeled by the relative boson number $\Delta N_L=N_L-N_L^0$ of left half cylinder in each tower
 ($N_L^0$ is the boson number of the state of reduced density matrix  with the largest eigenvalue).
 In each tower, the horizontal axis shows the relative momentum $\Delta K_y=K_y-K_y^0$
 in the transverse direction of the corresponding eigenvectors 
 ($K_y^0$ is momentum of the state with the largest eigenvalue in each tower).
 The numbers below the red dots label the nearly degenerating pattern for the low-lying ES with different  $\Delta K_y$.
 The black dashed line shows the entanglement gap in each momentum sector.} \label{ES}
\end{figure}

Firstly, we study the characteristic edge excitation with the help of entanglement spectrum (ES) \cite{Haldane2008}, 
as partitioning a cylinder into two halves manifests a ``spatial'' boundary.
Fig. \ref{ES} shows the ES for two of the groundstates $|\Psi^{L}_{\openone,s}\rangle$ with ``left'' chirality. 
The ES is grouped by the relative boson number $\Delta N_L$ of the half system
and their relative momentum quantum number $\Delta K_y$ (relative to the total $K_y$ of the highest weight spectrum level) 
along the transverse direction 
(referred to as y-direction).
The leading ES of $|\Psi^{L}_{\openone,s}\rangle$ displays 
the sequence of degeneracy pattern $\{1,1,2,3,5,...\}$ in each $\Delta N_L$ sector.
Importantly, the edge mode counting rule  agrees with the prediction of the free chiral boson  
in $SU(2)_1$ conformal field theory which describes the edge theory of Laughlin state \cite{Wen1990a}.
In addition, Fig. \ref{ES} also signals the chiral nature of the edge spectrum 
(ES increases as $K_y$ varies from $0$ to $-2\pi$ by a step $\delta K_y=-2\pi/L_y$), 
which results from the TRS spontaneously breaking. 
The other two groundstates $|\Psi^{R}_{\openone,s}\rangle$ with ``right'' chirality 
have opposite chirality ($\delta K_y=2\pi/L_y$) but the same degeneracy pattern in ES (not shown). 

\subsubsection{Fractional Charge and  Chern Number Quantization}
\begin{figure} 
 \begin{minipage}{0.98\linewidth}
 \centering
  \includegraphics[width=0.45\linewidth]{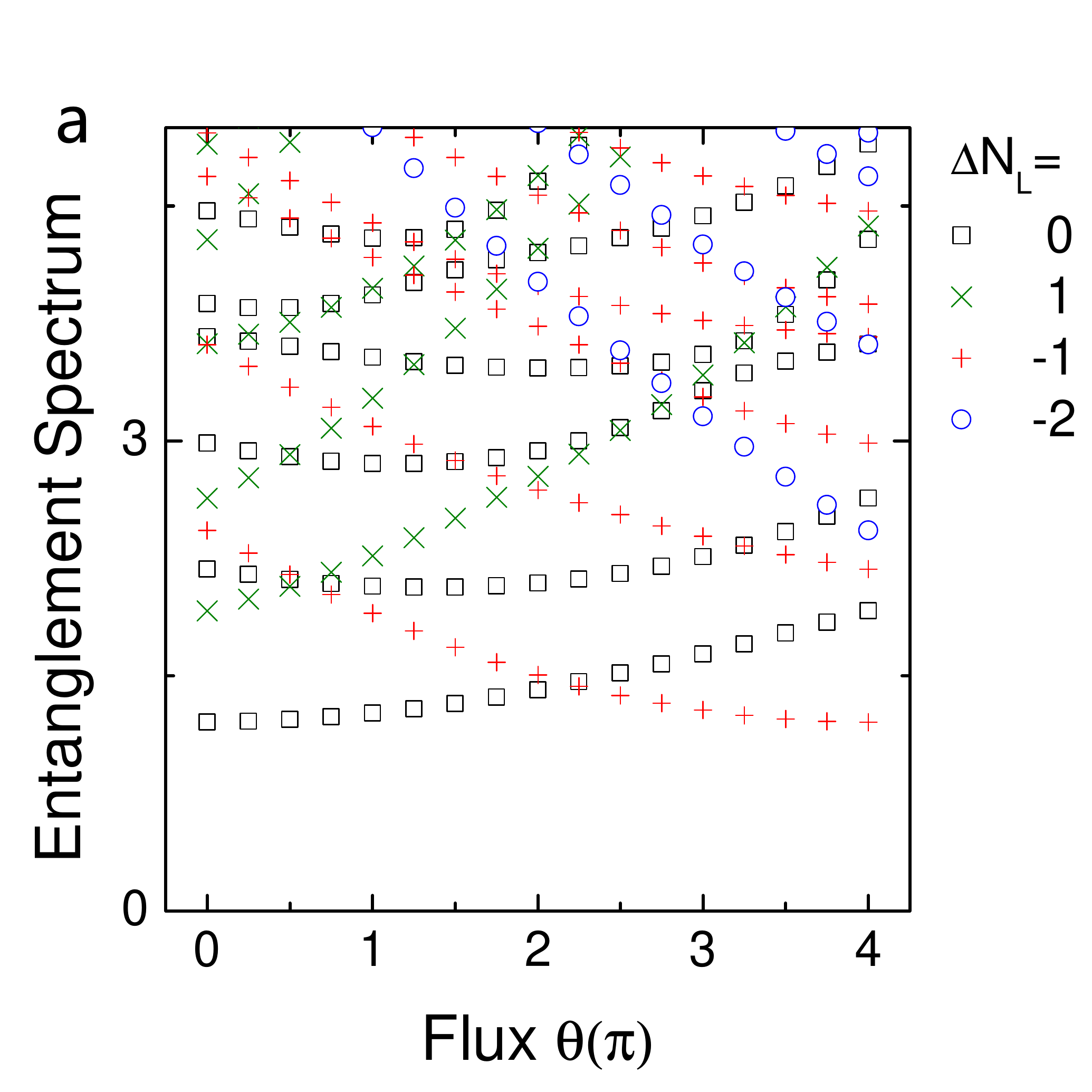}
  \includegraphics[width=0.45\linewidth]{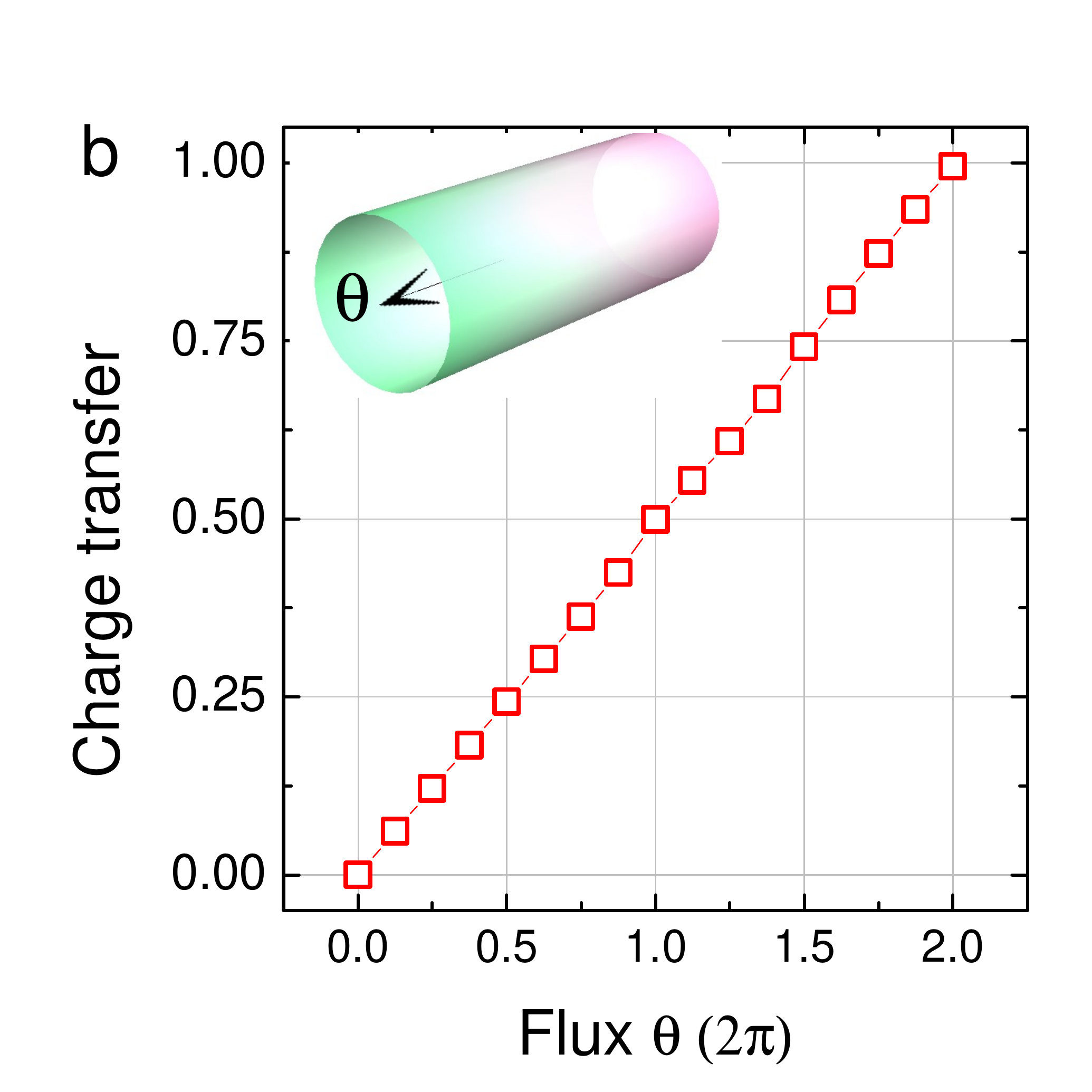}
 \end{minipage}
 \caption{(a) The ES flow with inserting flux $\theta$ in the hole of the cylinder with $L_y=4$.
 Starting from the identity ground state $|\Psi^L_{\openone}\rangle$ at $\theta=0$,
 the system evolves into the ground state $|\Psi^L_{s}\rangle$ by adiabatically threading a $\theta=2\pi$ flux.
 Further increasing flux up to $4\pi$ will drive the system  back to the identity ground state $|\Psi^L_{\openone}\rangle$.
 (b) Charge transfer from one edge to the other edge on the cylinder geometry.} \label{flux}
\end{figure}

Secondly, we perform a numerical flux insertion simulation on cylinder systems  \cite{SSGong_SR,WZhu_JSM,Zaletel_JSM,YCHe2014_PRB},
to determine the quantization of Hall transport and the topological Chern number of the ground state.
This simulation realizes  Laughlin gedanken experiment \cite{Laughlin1981,Oshikawa2000,Sheng2003},
where a quantized charge will be pumped from one  edge to the other  edge by inserting a $U(1)$ charge flux in the hole of the cylinder
for a quantum Hall state.
As shown in Fig. \ref{flux} (a), by threading a flux quantum $\theta=2\pi$, 
$|\Psi^L_{\openone}\rangle$ adiabatically evolves into $|\Psi^L_{s}\rangle$.
Further increasing flux up to $4\pi$ will drive the system  back to the $|\Psi^L_{\openone}\rangle$.
Interestingly, comparing the ES at $\theta=0$ and $4\pi$, 
the adiabatic flux insertion shifts the lowest level of ES from $\Delta N_L=0$ to $\Delta N_L=-1$,
signaling a net charage transfer $\Delta Q=1$ (a unit charge) from the left edge to the right edge.
In Fig. \ref{flux}(b),  the net  charge transfer $\Delta Q$ is demonstrated, 
which  is nearly quantized 
at $\Delta Q\approx 0.50$ at $\theta=2\pi$.  
Based on these observations, we identify the bulk Chern
number of the groundstate as $C_{\openone}=C_{s}= 1/2$, fully characterizing the
obtained state as the Laughlin $\nu=1/2$ state.

Here we identify each groundstate hosts a fractional Chern number $C=1/2$, 
which is similar to the Laughlin $\nu=1/2$ state in fractional Chern insulator \cite{YFWang2011}.
The key difference is, the ground state in the fractional Chern insulator
 inherits the non-trivial topology from the non-interacting band. 
However, our extended Bose-Hubbard model  (Eq. \ref{hamilton}) 
has a Chern number $C=0$ for the single-particle band.  
Thus, our  model realizes an interaction-driven topological phase 
from a topologically trivial band structure \cite{Simon2015}
resulting from the  spontaneous TSR breaking.

\subsubsection{Modular Matrix}
We further demonstrate the fractional quasiparticles in the bulk satisfy the so-called ``semionic'' statistics, 
as expected for Laughlin $\nu=1/2$ state. 
In the topological quantum field theory,
quasiparticle statistics is encoded in the modular matrices which describe the action of modular transformation
on the topological groundstates \cite{Wen1990b,CFT_Book,Dong2008}.
The $\mathcal{S}-$matrix contains the mutual statistics information of the
anyonic quasiparticles, such as quantum dimensions and fusion rules between different quasiparticles.
The $\mathcal{U}-$matrix encodes  the self-statistics of the quasiparticles, i.e. topological spin $h_i$.
Here we utilize the route of the ``twist'' overlap between the two topologically degenerating groundstates 
$|\Psi^L_{\openone}\rangle$ and $|\Psi^{L}_{s}\rangle$ 
to construct the modular $\mathcal{S}$ and $\mathcal{U}$ matrices \cite{YZhang2012,Cincio2013,WZhu2013}.

The obtained results at $V_1=0.5,V_2=V_3=1.0$ on $L_y=4$ cylinder are
\begin{eqnarray*}
&&\mathcal{S} \approx
\frac{1}{\sqrt{2}}\left[\begin{array}{cc}
1 & 1 \\
1 & -1
\end{array}\right]+
\left[\begin{array}{cc}
0.029 & 0.053 \\
0.005 & 0.036+0.033i
\end{array}\right] \\
&&\mathcal{U} \approx e^{-i \frac{3\pi}{4}}
\left[\begin{array}{cc}
1 & 0 \\
0 & i
\end{array}\right] \times
\left[\begin{array}{cc}
1 & 0 \\
0 & e^{i0.03\pi}
\end{array}\right].
\end{eqnarray*}
Indeed, the numerical obtained modular matrices are very  close to the analytical prediction
from $SU(2)_1$ Chern-Simons theory  \cite{Wen1990b,Dong2008,CFT_Book}: 
$\mathcal{S}^{SU(2)_1} =
\frac{1}{\sqrt{2}}\left[\begin{array}{cc}
1 & 1 \\
1 & -1
\end{array}\right]$ and
$\mathcal{U}^{SU(2)_1} = e^{-i \frac{3\pi}{4}}
\left[\begin{array}{cc}
1 & 0 \\
0 & i
\end{array}\right]$.
From $\mathcal{S}-$ and $\mathcal{U}-$ matrices, we obtain 
full statistics of fractionalized quasiparticles: 
1) There are two kinds of quasiparticles in total: 
Identity $\openone$ and semion $s$, and the total quantum dimension is $\mathcal{D}=\sqrt{2}$;
2) The fusion rule of quasiparticles (that specifies how the quasiparticles combine and fuse) \cite{CFT_Book}: 
$\openone\times s=s$, $s\times s=\openone$;
3) $\openone$ and $s$ respectively has topological spin (the phase factor for the quasiparticle obtained during a self-rotation of $2\pi$): 
$h_{\openone}=0$, $h_{s}=1/4$.
These braiding statistics provides the strongest confirmation that the topological groundstate is equivalent to
the bosonic FQH $\nu=1/2$ state.

\section{Discussion and Conclusion}
We have established a phase diagram for the Bose-Hubbard model for kagome system at 1/3 filling number
with the emergent FQH phase in the intermediate regime. 
We also address the nature of the quantum phase transitions between the FQH phase and 
solid phases (see Appendix \ref{QPT}). We utilize several quantities, such as entanglement entropy, 
correlation length and groundstate wavefunction fidelity.
The numerical evidences signal the first order character of the phase transition between
FQH phase and the strip phase as well as the charge density wave state. 
Moreover, we also find that in all three phases,  the obtained correlation length is much smaller
than the cylinder width $L_y$,
which  confirms that our DMRG calculation offers a reliable phase diagram  for the thermodynamic limit.

Regarding the laboratory realization for the emergent FQH state, 
a natural experimental setting for Bose-Hubbard physics is ultracold atomic gases \cite{Bloch2008}.
One advantange of our model is that it only contains  real nearest-neighbor hoppings 
and density-density interactions.
Moreover, we also note that
there are several existing candidates of spin$-1/2$ materials with kagome structure, such as
$BaCu_3V_2O_8(OH)_2$\cite{Vesignieite}, $Cu_3(Mg,Zn)O_7(OH)_2\cdot H_2O$ \cite{Janson2008}, 
$Cu_3V_2O_7(OH)_2\cdot H_2O$\cite{Ishikawa2015}, 
$Rb_2Cu_3SnF_{12}$\cite{deformedkagome} and $Dy_3 Ru_4 Al_{12}$ \cite{Gorbunov2014}, 
and each of them  has its own interactions deserving  to be studies more carefully
under magnetic field for possible detecting of the exotic magnetization plateaus. 
But the fine tuning of second and third nearest neighbor interactions in such materials may be difficult to achieve.
So the material-realization of our proposed FQH phase in condensed matter setting
may depend on  synthesising  more kagome materials in the future.

In conclusion, we have presented a global phase diagram of an extended Bose-Hubbard model 
on the kagome lattice at fractional one-third filling.
Importantly, the interplay between the underlying lattice and strong interaction 
gives birth to a fractional quantum Hall (FQH) liquid phase,
even though the non-interacting band structure is topological trivial.
The FQH phase we present here provides a ``proof of the principle''  example of
interaction-driven quantum anomalous Hall effect with 
time-reversal symmetry (TRS) spontaneously breaking.
We also provide complete characterization of the universal properties
of the FQH phase, including ground state degeneracy, 
topological entanglement  spectrum, fractionally quantized  Chern number 
and anyonic quasiparticle statistics.
To our best knowledge, this is the first example of 
a TRS breaking topological phase at one-third filling on kagome lattice system.

We believe our current work will inspire upcoming research efforts both 
in theoritical  and experimental fields. 
From the theoretical side,
the present calculations show that the model of strongly interacting 
hard-core bosons can harbor rich and interesting phases through interaction engineering.
It is also interesting to study the interacting 
particles to be spinless or spinful fermions, since the possible topological liquid phase
in fermionic models has been sought for a long time \cite{Raghu2008,WenJun2010}.
From the experimental side, our work will
provide direction and insight  in searching for the topological liquid phase 
in realistic materials with kagome lattice structure, or by
engineering such systems in ultracold atomic settings.
Thus our present findings would provide both theoristes and experimentalists a rich playground 
in searching  of new topological phases induced by strong interaction.  

\textit{Acknowledgements.---}
We thank Y. Zhang for stimulating discussions.
This research is supported by 
the U.S.  Department of Energy, Office of Basic Energy Sciences 
under grants No. DE-FG02-06ER46305 (W.Z., D.N.S).
S.S.G is also supported by the National High Magnetic Field Laboratory 
(NSF DMR-1157490) and the State of Florida.

%

\vspace{40pt}
\begin{appendix}



\section{Method} \label{method}
In this paper, the calculations are based on the density-matrix renomralization group (DMRG) algorithm on cylinder geometry \cite{SWhite1992,McCulloch2008}
and the exact diagonalization (ED) on torus geometry, both of which have been proven to be effective
and complementary  tools for studying realistic models containing arbitrary strong and frustrated interactions.  
On one hand, ED is staightforward in identifying   the groundstate degeneracy on compactified spaces. But the drawback is that with the exponential growing
of the Hilbert space, the accesible systems are limited to smaller sizes,  up to $N_s=36$ for this study. On the other hand, DMRG calcualtion allows us to obtain accurate groundstates and related entanglement measurements on much larger system sizes beyond the ED limit.
Moreover, DMRG calculation also has the  advantages  
of   probing ground states with spontaneous symmetry breaking and topological ordering. 
The DMRG calculations on long cylinder tend to automatically select the groundstates with minimal entropy \cite{HCJiang2012}, which is helpful to study the fractionalized quasiparticle statistics in topological ordered states \cite{Cincio2013}.



\subsection{Details of DMRG Calculation}
We study the cylinder system with open boundaries in the x direction
and periodic boundary condition in the y direction. 
The available system sizes are cylinders of circumference $L_y=3,4,5,6$ (in unit of unit cell).
For the largest system width ($L_y = 6$), we keep up to $M=8400$ $U(1)$
states and reach the DMRG truncation error $5 \times 10^{-7}$. 

The entanglement entropy and spectrum can be easily obtained in the DMRG.
By partitioning the system into subsystems A and B, the groundstate wavefunction $|\psi\rangle$ 
can be decomposed according to Schmidt decomposition $
|\psi\rangle=\sum_i \lambda_i^{1/2} |\psi^i_A\rangle  |\psi^i_B\rangle$,
where $\lambda_i$ are eigenvalues of 
the reduced density matrix $\hat{\rho}_A$ of subsystem A.
Thus the entanglement entropy can be defined as $
S_A=-tr[\hat{\rho}_A \ln \hat{\rho}_A] = -\sum_i \lambda_i\ln \lambda_i $.
The eigenvalues $log\{\lambda_i\}$  plotted against the relative momentum quantum number  $\Delta k_y$ of the subsystem A,
is defined as the entanglement spectrum \cite{Haldane2008}.

\subsection{Adiabatic DMRG and Fractionally Quantized Chern Number}
We have used the numerical flux insertion experiment based on the adiabatical
DMRG simulation to detect the topological Chern number
of the bulk system \cite{SSGong_SR,WZhu_JSM,Zaletel_JSM}. 
To simulate the flux $\theta$ threading in the hole of a cylinder, we impose the twist boundary conditions along 
the y direction with replacing terms $b^{\dagger}_{\mathbf{r}^{\prime}}b_{\mathbf{r}}+h.c.\rightarrow
e^{i\theta_{\mathbf{r}^{\prime}\mathbf{r}}}b^{\dagger}_{\mathbf{r}^{\prime}}b_{\mathbf{r}}+h.c.$ 
for all neighboring $(\mathbf{r}, \mathbf{r}^{\prime})$ bonds
with hoppings crossing the y-boundary in the Hamiltonian (Eq. \ref{hamilton}).
The charge pumping from one edge to the other edge can be computed from
$\langle \Delta Q(\theta)\rangle=Tr[\hat{\rho}_L(\theta) \hat{Q}(\theta)]$, where 
$\hat{Q}(\theta)$ is the $U(1)$ quantum number and $\hat{\rho}_L(\theta)$ is reduced density matrix of left half system.
Due to the quantized Hall response, the Chern number of ground state is equal to 
the charge pumping by threading a $\theta=2\pi$ flux \cite{Laughlin1981}.
To realize the adiabatic flux insertion, we use the step of flux insertion as $\Delta \theta=0.25\pi$.

\section{Classical Phase Diagram} \label{classic}
\begin{figure}[!htb]
  \includegraphics[width=0.48\linewidth]{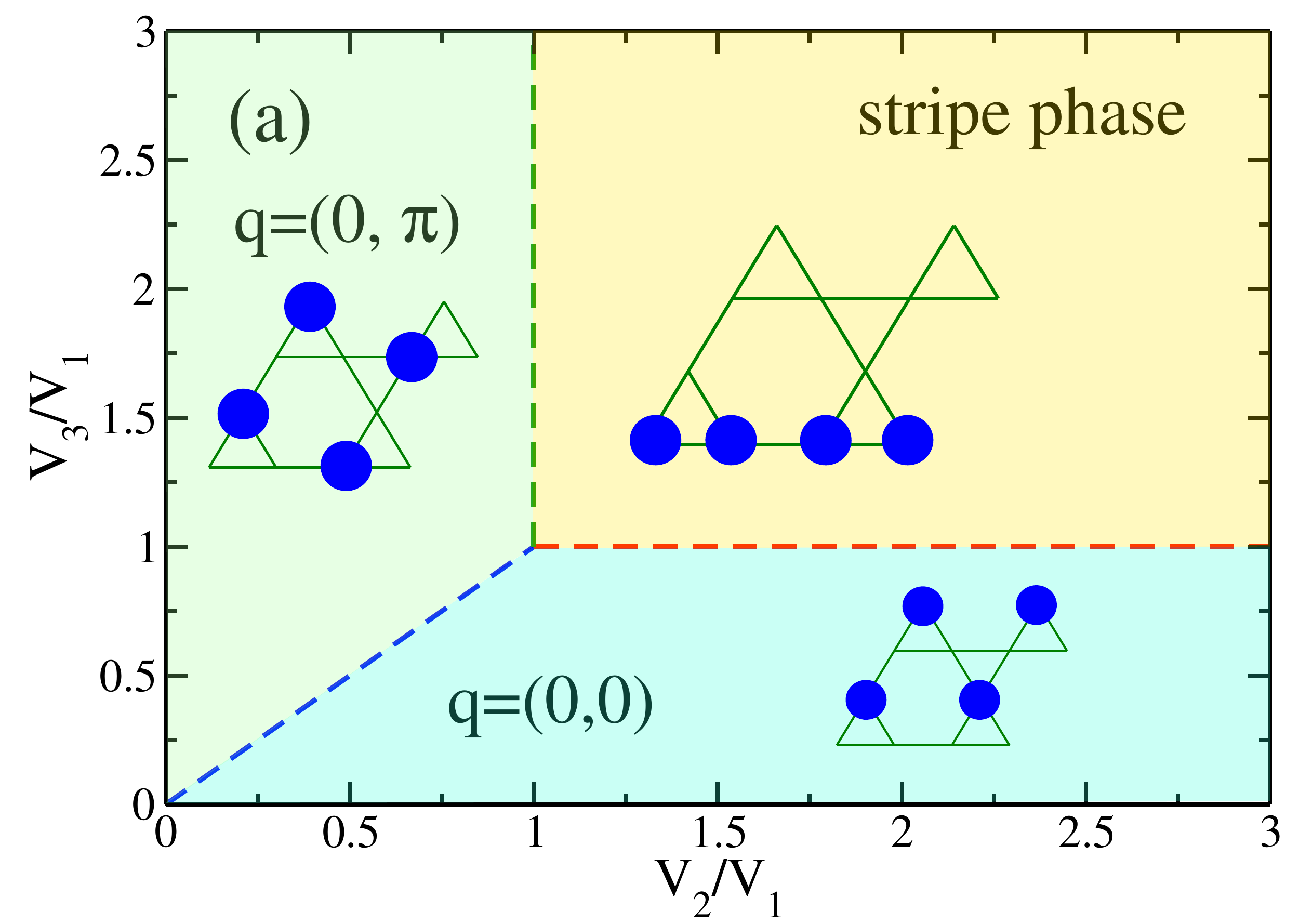}
  \includegraphics[width=0.48\linewidth]{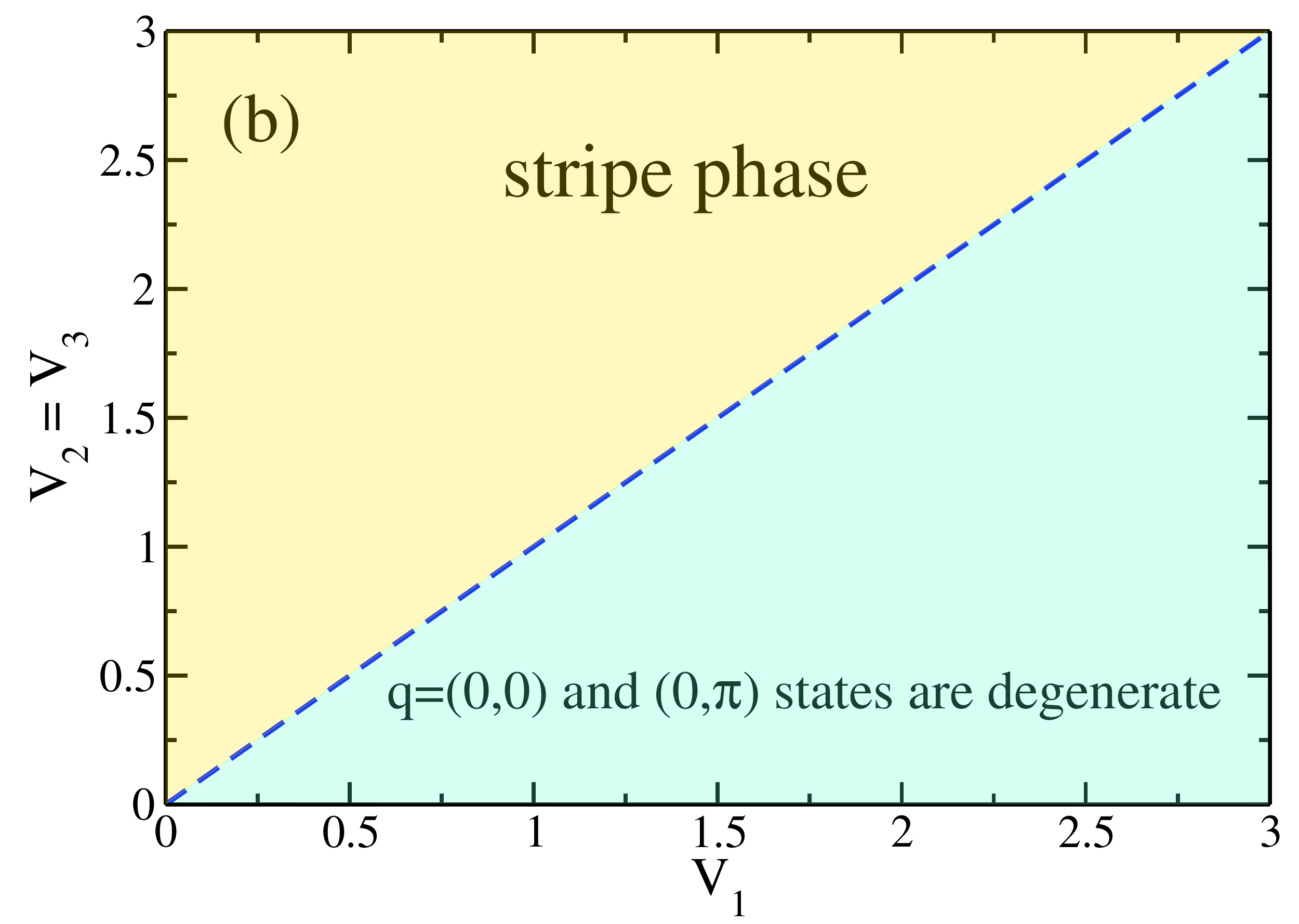}
 \caption{Classical phase diagram in $\{V_1,V_2,V_3\}$ parameter space.
 (a)The phase diagram plotted in $V_2/V_1$ and $V_3/V_1$ parameter space. The insets show the unit cells of the different phases, where the large and small circles denote the occupied and the unoccupied sites. (b)The phase diagram plotted in $V_1$ and $V_2=V_3$ parameter space.} \label{classical} 
\end{figure}

In this section, we discuss the phase diagram of the model without hopping term $t=0$,
where the system reduces to a classical Ising model with competing antiferromagnetic
interactions \cite{Capponi2015}. We compare the energy of four states, i.e., the $q=(0,0)$ state, the $q=(0,\pi)$ state,
the stripe state, and the $\sqrt{3}\times \sqrt{3}$ state. The pattern of these states are shown in
the inset of Fig.~\ref{classical}(a). Interestingly, we find that when both $V_2, V_3$ are larger
than $V_1$, the stripe state always has the lowest energy; otherwise, the $q=(0,0)$ and $q=(0,\pi)$
states compete depending on the strengths of $V_2$ and $V_3$ (see Fig.~\ref{classical}(a)).

It is interesting to compare this classical phase diagram with the quantum phase diagram as shown in Fig.~\ref{phase}(a).
To compare them, we set $V_2=V_3$ as shown in Fig.~\ref{classical}(b), where
the system shows two phase regions, a stripe phase and a charge density wave phase region
with the boundary at $V_2=V_3=V_1$. Interestingly, the charge density wave states with
$q=(0,0)$ and $q=(0,\pi)$ are degenerated in the case of $V_1 < V_2$. In the quantum case,
the degeneracy is lifted with the $q=(0,0)$ state having the lower energy. The FQH 
phase obtained by DMRG appears in the transition region between the stripe phase and the $q=(0,0)$ phase
of the classical phase diagram. 
The emergence of the topological phase appears seems to arise as a result  
that quantum fluctuations destory long-ranged orders around the transition region.
This could serve as a guiding principle for finding topological phases in other models and/or on other lattices.

\begin{figure}[!htb]
 \includegraphics[width=0.75\linewidth]{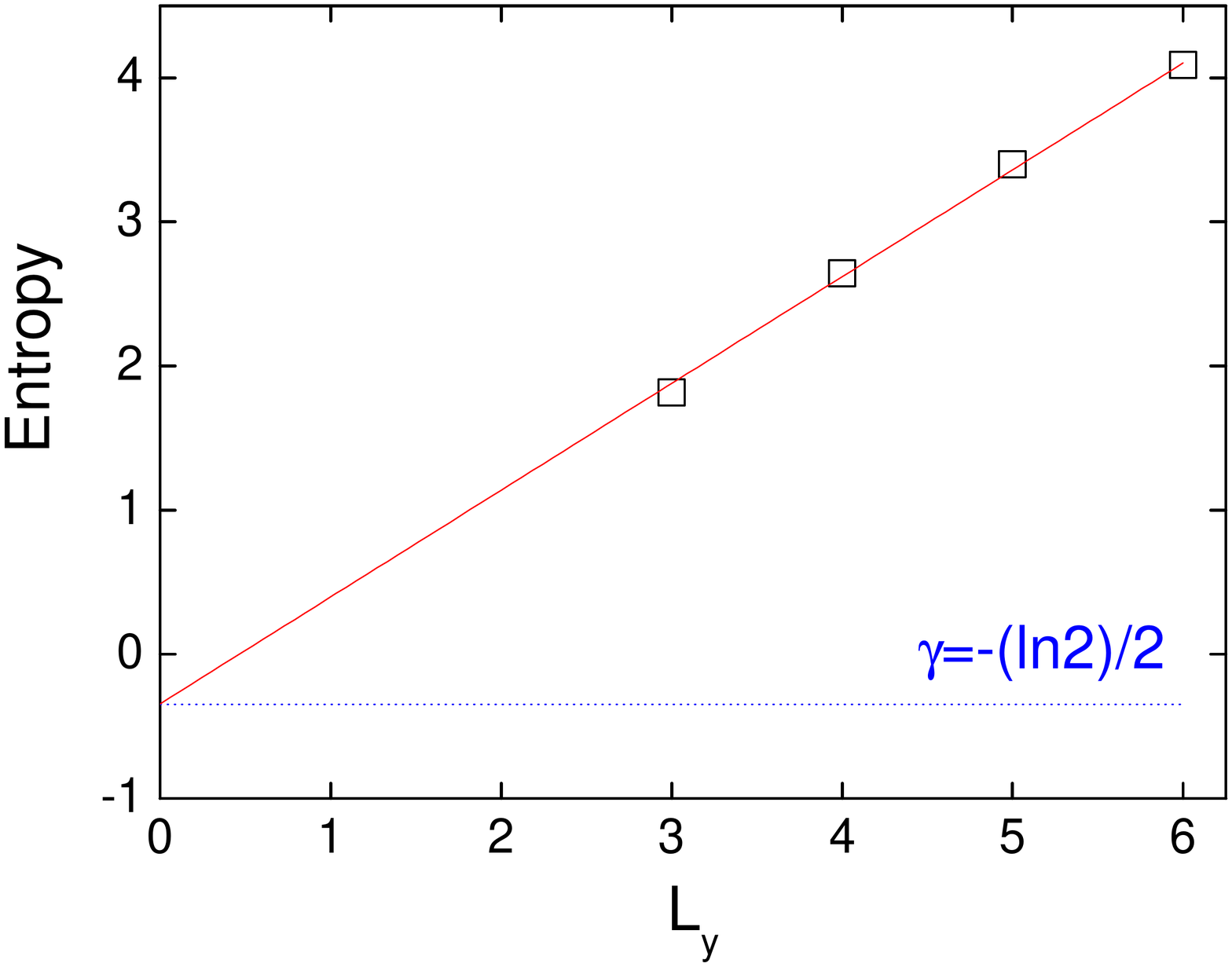}
 \caption{Entanglement entropy scaling with $L_y$.} 
\end{figure}

\section{Topological entanglement entropy} \label{TEE}
For a gapped quantum phase with topological order, 
the topological entanglement entropy (TEE) $\gamma$ is proposed to characterize
the non-local entanglement \cite{Kitaev2006,Levin2006}. Generally speaking, entropy has the form $S = \alpha L_y - \gamma$,
where $L_y$ is the boundary of the subsystem, and $\alpha$ is a non-universal constant.
While a positive $\gamma$ is a correction to the area law of entanglement and 
reaches a universal value determined by total quantum dimension $\mathcal{D}$ of quasiparticle excitations
as $\gamma = \ln \mathcal{D}$. 
For the $\nu = 1/2$ Laughlin state, the quantum dimension of each quasiparticle is $1$ (see main text), leading to 
the total quantum dimension $\mathcal{D} = \sqrt{2}$ and thus the TEE $\gamma =\ln \mathcal{D}= \frac{1}{2}\ln 2$.

By using the DMRG simulations, 
we obtain the minimal entropy state \cite{HCJiang2012,YZhang2012} with spontaneously broken time-reversal
symmetry and calculate the corresponding Von Neuman entanglement entropy. The converged entropy 
are available for $L_y = 3,4,5,6$ cylinders.  
For $V_1=0.5,V_2=1.0,V_3=1.0$, we make a linear fitting of the
entropy data for $L_y =3,4,5,6$ cylinders, and find the TEE $\gamma \approx 0.343 \pm 0.075$. 
(If we make the linear fitting based on data for $L_y=4,5,6$ (not shown), the obtained result is $\gamma \approx 0.241 \pm 0.098$).
Despite some uncertainty in the  fitting, the obtained TEE approaches the prediction of the $\nu = 1/2$
Laughlin state $\gamma = \frac{1}{2}\ln 2 \approx 0.346$.

\begin{figure} 
 \includegraphics[width=0.65\linewidth]{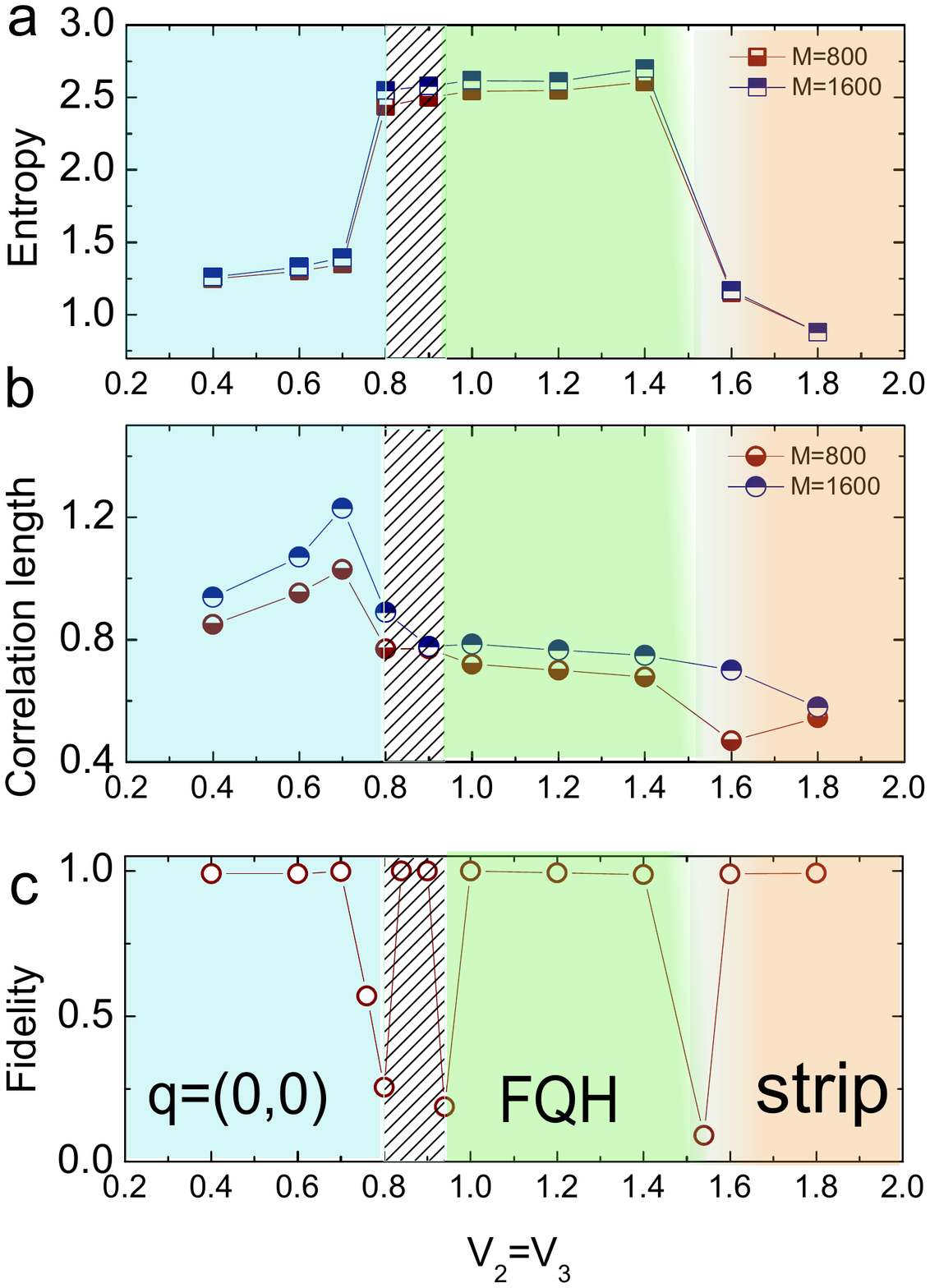}
 \caption{(a) Entanglement entropy and (b) correlation length versus $V_2(=V_3)$ by setting $V_1=0.5$. The calculations are performed on $L_y=4$ cylinder, by keeping $M=800$ (brown dots) and $1600$ (blue dots) states. (c) Wavefunction fidelity plotted as $V_2(=V_3)$ by setting $V_1=0.5$. The calculations are performed on $L_y=4$ cylinder, by keeping $M=800$.} \label{transition}
\end{figure}

\section{Quantum phase transitions} \label{QPT}
In order to uncover the nature of correponding phase transitions between
FQH phase and solid phases, we
inspect several quantities that are expected to be sensitive to a phase transition,
such as the entanglement entropy $S$, and correlation length $\xi$.
Both quantities are expected to show a finite jump when crossing a first order transition.
We also calculate the groundstate wavefunction fidelity $F=|\langle\psi(V)|\psi(V+\delta V)\rangle|$ 
($V$ is some parameter in Hamiltonian) \cite{McCulloch2008}, 
which can faithfully describe the first-order transition or energy level crossing.

We show the results along the reference line in Fig. 1 (in the main text), 
by fixing $V_1=0.5$ and varying $V_2=V_3$.
In Fig. \ref{transition}(a), it is found that the entanglement entropy shows a sharp jump around 
$V_2=V_3\approx 0.8$ and a drop around $V_2=V_3\approx 1.5$. 
We also observe the similar behavior when looking at the correlation length, as shown in Fig. \ref{transition}(b).
Both of these two measurements signals a direct first order phase transition between
FQH phase and solid phases.
In addition, we also find that, between the FQH phase and charge density wave $q=(0,0)$ phase, 
there exists a narrow window for the coexistence of both FQH nature and the charged order, 
as shown by the shaded area in the phase diagram. We have checked that, in this intermediate regime,
the groundstate hosts the quantized Chern number $C=1/2$, but develops weak charge order. From the 
wavefunction fidelity in Fig. \ref{transition}(c), 
it is shown a first-order transition between the coexistence region and the FQH phase ($q=(0,0)$ phase).


Moreover, when studying the topological order on cylinder geometry with finite width $L_y$, 
the correlation length $\xi$ of the ground state offers a natural consistency check for the assumption 
that the value of $L_y$ is large enough to be representative of the thermodynamic limit. 
The correlation length is defined by $\xi=-\ln |\epsilon_1/\epsilon_2|$, where $\epsilon_{1,2}$ 
are two largest eigenvalue from transfer matrix \cite{McCulloch2008}.
If $L_y$ is much larger than $\xi$, we indeed expect finite size effects to be very small. 
Indeed, in Fig. \ref{transition}(b), the condition $L_y>\xi$ is satisfied so that our DMRG calculation offers a reliable and relevant results for thermodynamic limit. Based on these measurements, we expect the finite size effect should be small in our calculations.


\end{appendix}



\end{document}